\documentclass[prd,%
preprint,%
tightenlines,%
showpacs,%
nofootinbib,%
superscriptaddress,%
amsfonts,%
]{revtex4-1}

\begin{document}

\begin{flushright}
UFIFT-QG-11-02
\end{flushright}

\title{Nonlocal metric formulations of MOND with sufficient
lensing}

\author{C\'edric~Deffayet} \email{deffayet@iap.fr}
\affiliation{AstroParticule \& Cosmologie, UMR 7164-CNRS,
Universit\'e Denis Diderot-Paris 7, CEA, Observatoire de Paris, 10
rue Alice Domon et L\'eonie Duquet, F-75205 Paris Cedex 13, France}
\affiliation{${\mathcal{G}}{\mathbb{R}}
\varepsilon{\mathbb{C}}{\mathcal{O}}$, Institut d'Astrophysique de
Paris, UMR 7095-CNRS, Universit\'e Pierre et Marie Curie-Paris 6,
98bis boulevard Arago, F-75014 Paris, France}

\author{Gilles~\surname{Esposito-Far\`ese}} \email{gef@iap.fr}
\affiliation{${\mathcal{G}}{\mathbb{R}}
\varepsilon{\mathbb{C}}{\mathcal{O}}$, Institut d'Astrophysique de
Paris, UMR 7095-CNRS, Universit\'e Pierre et Marie Curie-Paris 6,
98bis boulevard Arago, F-75014 Paris, France}

\author{Richard P. Woodard} \email{woodard@phys.ufl.edu}
\affiliation{Department of Physics, University of Florida, FL-32611
Gainesville, USA}

\begin{abstract}
We demonstrate how to construct purely metric modifications of
gravity which agree with general relativity in the weak field
regime appropriate to the solar system, but which possess an
ultra-weak field regime when the gravitational acceleration
becomes comparable to $a_0 \sim 10^{-10}~{\rm m/s}^2$. In this
ultra-weak field regime, the models reproduce the MOND force without dark matter and also give enough gravitational
lensing to be consistent with existing data. Our models are
nonlocal and might conceivably derive from quantum corrections to
the effective field equations.
\end{abstract}


\maketitle

\section{Introduction}
Although Einstein's gravitational field equations are in
remarkable agreement with all solar-system and binary-pulsar
tests \cite{GR}, they lead to a cosmological model which needs
that the energy density of the Universe is strongly dominated by
components, dark matter and dark energy, which have so far eluded
direct detection. In particular, there is not enough baryonic
matter to explain the observed properties of galactic dynamics
using standard Einsteinian (or actually Newtonian) gravitational
equations. The usual solution to this problem is to suppose that
the vast majority of nonrelativistic matter in the universe
consists of some weakly interacting particle we have not yet
detected. Although there are several reasonable candidates for
what this dark matter might be (see, e.g., \cite{DMreview}), it
is worthwhile considering the alternative: It is possible that
the field equations break down at galactic scales, i.e., that
gravity is modified at distances relevant for dealing with
galactic and inter-galactic dynamics.

Along this line, Milgrom proposed a simple phenomenological law
\cite{Milgrom1} which leads to successful explanations of various
observations, and to predictions \cite{Milgrom3} which turned out
to be confirmed {\it a posteriori}. Milgrom's proposal, Modified
Newtonian Dynamics (MOND), stipulates that a test particle at a
distance $r$ from a mass $M$ will experience a gravitational
acceleration given by the Newtonian expression $a_N = GM/r^2$ as
long as $a_N$ is (much) larger than a critical acceleration
$a_0$, while the same particle will undergo the MOND acceleration
$a_{\text{MOND}} = \sqrt{a_N a_0} = \sqrt{GM a_0}/r$ when $a_N$
is  smaller than $a_0$. (Although Milgrom's proposal can be
viewed as a change in Newton's 3rd law \cite{Milgrom2} it is more
often imagined as a modification of gravity, which is the view we
shall take.) It turns out that the value \cite{BBS}
\begin{equation}
a_0 \approx 1.2 \times 10^{-10}\, \text{m}.\text{s}^{-2} \label{a0}
\end{equation}
allows an excellent fit of galaxy rotation curves using
reasonable mass-to-luminosity ratios \cite{BdJ}, without the need
for non-baryonic dark matter \cite{SanMc}. MOND's ability to
explain certain observed regularities of galactic structure
contrasts favorably with dark matter, for which these
regularities must either be accidental or else the result of some
yet-to-be-discovered attractor solution in structure formation.
Indeed, in the case of rotationally supported systems, MOND
provides a simple explanation for (i)~the Tully-Fisher relation
\cite{TF}, which states that the observed limiting rotation
velocity of galaxies, $v_{\infty}$, scales as the fourth root of
the baryonic mass of the galaxy (see \cite{Stacy} for a recent
dramatic confirmation of this relation); (ii)~Milgrom's law,
stating that the need for dark matter always seems to occur when
the gravitational acceleration falls to about $a_0$
\cite{Kaplinghat:2001me}; (iii)~Freeman's law, namely that the
surface density never exceeds $a_0/G$ \cite{Sanders:2008iy}; and
(iv)~Sancisi's law, i.e., that bumps in the rotation curves are
correlated to the baryonic mass \cite{Sanders:2008iy}. For
pressure-supported systems, MOND also explains their typical size
$R = \sqrt{G M/a_0}$ and predicts a stellar velocity dispersion
$\sigma \propto (G M a_0)^{Ê1/4}$
\cite{Sanders:2001ek,Sanders:2010aa}, explaining thus the
Faber-Jackson relation \cite{Faber:1976sn}. MOND also was able to
predict properties of low surface brightness galaxies that were
eventually confirmed by observation \cite{SMEB,EBSM}. Recently,
Ref.~\cite{Hernandez:2011uf} used a large catalog of widely
separated binary star systems \cite{Dhital:2010zv} as evidence for
the breakdown of Kepler's third law at the MOND acceleration
scale $a_0$.

On the other hand, the original MOND formulation does have some
difficulties. In particular, a single acceleration scale $a_0$
fitted to galaxy rotation curves does not account for velocity
dispersions in the cores of galactic clusters, which still need
some amount of dark matter \cite{ASQ}. Similarly, X-ray and weak
lensing data from the bullet cluster \cite{Clowe:2003tk} indicate
that dark matter exists at a different location from the gas.
(Because weak lensing data is involved, one must make some
assumption about how lensing occurs in MOND gravity, and this
might affect the negative conclusion
\cite{Dai:2008sf,Feix:2007zm,Angus:2007qj}.) A reasonable
fraction of dark baryons and/or massive neutrinos is thus still
required at cluster scales, even if the MOND scheme happens to be
an actual law of Nature at galaxy scales. This would not be in
contradiction with observation, since most of the baryons in the
Universe have not yet been detected. Dark baryons are actually
also required to explain the observed peaks in the CMB spectrum.

The most serious problem of MOND is that it is not a complete
theory, so that testing it often implies making guesses about its
predictions for lensing or for cosmological evolution. This has
not been for a lack of efforts, and a large number of theoretical
constructions have been proposed over almost three decades to
promote MOND to a consistent relativistic field theory. One major
problem has always been simultaneously reproducing the
Tully-Fisher relation and giving a sufficient amount of weak
lensing. That problem was finally surmounted in 2004 by the
tensor-vector-scalar (TeVeS) model constructed by Bekenstein
(after years of work with Milgrom and Sanders)
\cite{Bekenstein:1992pj,Bekenstein:1993fs,Sanders:1996wk,
TeVeS,Bekenstein:2004ca,Sanders:2005vd}, in which the MOND force
(implying the Tully-Fisher relation) is mediated by a scalar
field, and where the presence of a unit timelike vector field
helps in obtaining the right amount of light deflection from
galaxies and clusters. The model has been shown to give better
agreement with cosmological data than many believed any
relativistic extension of MOND could do
\cite{Skordis,Skordis2,Dodelson,Bourliot,ZFS,Skordis:2009bf,
Clifton:2011jh}.

In its original formulation \cite{TeVeS}, TeVeS suffers from several
theoretical and experimental difficulties
\cite{Bruneton:2007si,Ferreras:2009rv,Mavromatos:2009xh}, notably a
serious instability \cite{Clayton:2001vy,Seifert}. The latest
version of TeVeS \cite{Sagi:2009kd,Bekenstein:2010pt}, inspired by
the Einstein-Aether framework
\cite{Jacobson:2000xp,Jacobson:2004ts,Eling:2004dk,Foster:2005dk,Jacobson:2008aj,refB},
seems to avoid this instability and predicts post-Newtonian
parameters consistent with solar-system tests. However, it still
needs an unnaturally fine-tuned function of the scalar kinetic term
in its action to be also consistent with binary-pulsar tests
\cite{Bruneton:2007si}. (The extended Vainshtein mechanism recently
proposed in \cite{Babichev:2011kq} is a way to avoid this
difficulty.) Out of the many alternative models which have been
proposed in the literature (for example, see \cite{Li:2009zzh}), the
recent bi-metric theory \cite{Milgrom:2009gv} is a particularly
promising and elegant one, although its detailed properties (notably
its stability) remain to be fully understood.

As promising as we consider TeVeS to be, its dependence on other
fields to carry part of the gravitational force is somewhat
counter to the spirit of relativity. In the present paper, we
re-examine pure-metric formulations of MOND along the general
lines previously considered in \cite{SW2}.\footnote{By
``pure-metric'', we mean that the full gravitational interaction
is described by the dynamics of a single metric tensor
$g_{\mu\nu}$, without introducing explicit extra fields like
scalars or vectors, although such degrees of freedom may actually
be hidden in some excitations of $g_{\mu\nu}$. For instance, we
would call pure-metric the class of $f(R)$ models, although it is
well known they are equivalent to specific scalar-tensor
theories. Our phrase pure-metric should also be distinguished
from what is called a ``metric theory'' in \cite{GR}, meaning
there that matter is minimally coupled to a single metric tensor
$g_{\mu\nu}$. What we call pure-metric is a subclass of such
metric theories, but we also impose that the kinetic term of
gravity itself is a functional of only $g_{\mu\nu}$.} Our aim is
not to here produce the ultimate theory but rather just to show
what form any pure metric generalization of MOND must take in
order to combine two key features for a static, spherically
symmetric and pressureless source which contains no dark matter:
\begin{itemize}
\item{reproduce the MOND force law in the ultra-weak field regime
of accelerations comparable to $a_0$; and}
\item{produce enough weak lensing to be compatible with
observations.}
\end{itemize}
We first derive the form the MOND corrections to the Lagrangian
must take in order to combine these properties when specialized
to a static and spherically symmetric geometry. Then we
demonstrate that no local curvature scalar has this form.
However, nonlocal scalars do exist which take the correct form,
and we exhibit some. As anticipated in \cite{SW2}, the Lagrangian
inevitably becomes cubic in the weak fields, raising concerns
about stability which we discuss briefly in the conclusion.

This paper is organized as follows. In Sec.~\ref{PHENO}, we
discuss some basic phenomenological properties used to deal with
static, spherically symmetric systems, and we show how a pure
metric action reproducing the MOND phenomenology can be devised for such systems.
In Sec.~\ref{LOCAL}, we demonstrate that such an action cannot be
local, i.e., it cannot be a function of only the metric and a
finite number of its derivatives. Section \ref{NONLOCAL}
introduces the main ingredients needed to construct a suitable
nonlocal action for gravity. In Sec.~\ref{MODEL} we exhibit a
nonlocal model having the properties discovered in
Sec.~\ref{PHENO}, i.e., which reproduces the MOND dynamics at
large distances, including enough weak lensing, while tending
towards general relativity at small distances. Our conclusions
are given in Sec.~\ref{DISCUSSION}.

\section{Phenomenology}\label{PHENO}

The point of this section is to derive the form that the MOND
modification to the gravitational Lagrangian must take when
specialized to the ultra-weak field regime of a static,
spherically symmetric geometry,
\begin{equation}
ds^2 = -B(r) c^2 dt^2 + A(r) dr^2 + r^2 d\Omega^2 \; . \label{SSPHE}
\end{equation}
We begin by reviewing how the equations of general relativity work
for a source which would, of course, need to consist mostly of dark
matter. In the ultra-weak field limit, these equations imply
relations for the two linearized potentials, $a(r) \equiv A(r) - 1$
and $b(r) \equiv B(r) - 1$. One of these relations determines how
the potentials depend upon the source and the other fixes how they
depend upon each other. Our metric interpolation of MOND consists of
changing how the potentials depend upon the source but not much how they
depend upon one another. As the section closes we consider the form
the MOND correction to the gravitational Lagrangian must take in
order to substitute our MOND equations for those of general
relativity.

We assume a perfect fluid source,
\begin{equation}
T_{\mu}^{\nu} = {\rm diag}\left(-\rho, P,P,P \right),
\end{equation}
where $\rho(r)$ and $P(r)$ are respectively the energy density and
pressure. Only two of the ten field equations are independent in
this geometry; the rest are either trivial or implied by
conservation. Defining $G$ as Newton's constant, the $tt$ and $rr$ Einstein equations
equations are,
\begin{eqnarray}
\frac{G_{tt}}{B} & = & \frac{A'}{r A^2} + \Bigl( \frac{A \!-\!
1}{r^2 A}\Bigr) = \frac{8\pi G \rho}{c^4} \; , \label{tteqn} \\
\frac{G_{rr}}{A} & = & \frac{B'}{r A B} - \Bigl( \frac{A \!-\!
1}{r^2 A}\Bigr) = \frac{8\pi G P}{c^4} \; . \label{rreqn}
\end{eqnarray}
Equation (\ref{tteqn}) can be integrated to give us the $rr$
component,
\begin{equation}
A(r) = \Bigl[1 \!-\! \frac{2 G M(r)}{c^2 r} \Bigr]^{-1} \; ,
\end{equation}
where the enclosed mass is
\begin{equation}
M(r) \equiv \frac{4\pi}{c^2} \int_0^r \!\! dr' \, {r'}^2 \rho(r').
\end{equation}
The second equation (\ref{rreqn}) could also be integrated but we
shall not need to do this.

Now consider the regime of zero pressure and very weak potentials,
for which the linearized potentials take the form,
\begin{equation}
a(r) \approx \frac{2 G M(r)}{c^2 r} \approx r b'(r).
\end{equation}
These relations can be expressed in many ways but a convenient form,
for our purposes, is as one equation for how the potentials depend
upon the source,
\begin{equation}
r b'(r) \approx \frac{2 G M(r)}{c^2 r} \; , \label{dmatter}
\end{equation}
and another equation for how the two potentials depend upon each
other,
\begin{equation}
a(r) \approx r b'(r). \label{lensing}
\end{equation}
The first equation (\ref{dmatter}) is what tells us that
explaining cosmic motions requires dark matter, whereas the
second equation (\ref{lensing}) tells us that the amount
of weak lensing is consistent with the data, assuming cosmic
motions are explained.

For circular geodesic motion at fixed radius $r$ with angular
velocity $\dot{\phi}$, one can show
\begin{equation}
r B'(r) = r b'(r) = \frac{2 r^2 \dot{\phi}^2}{c^2} = \frac{2 v^2}{
c^2} \; . \label{circular}
\end{equation}
We emphasize that relation (\ref{circular}) depends only upon the
geometry (\ref{SSPHE}) and minimal coupling to matter (that we will always assume within the present paper), without any assumption about the
gravitational field equations which produce it. The Tully-Fisher
relation implies that the rotational speed $v(r) = r \dot{\phi}$
tends to a constant which goes as the fourth root of the source
luminosity. MOND imposes the Tully-Fisher relation by changing
equation (\ref{dmatter}) to \cite{SanMc},
\begin{equation}
r b'(r) \longrightarrow \frac{2 \sqrt{a_0 G M(r)}}{c^2} \; ,
\label{MOND1}
\end{equation}
where the right arrow indicates that the relation applies in the
ultra-weak field regime of low accelerations.

Relation (\ref{MOND1}) contains the physics we want, but it is not
yet in the form of a modification to just the left hand side of
the gravitational field equations (\ref{tteqn}-\ref{rreqn}). To
reach that form we need to isolate the local energy density
$\rho(r)$ by first squaring, then differentiating and shifting
some factors from right to left,
\begin{equation}
\frac{c^2}{2 a_0 r^2} \Bigl( (r b')^2\Bigr)' = \frac{8\pi G
\rho}{c^4} \; . \label{MOND2}
\end{equation}
The other MOND equation can be written in a variety of ways
because the right hand side vanishes for the relevant case of
zero pressure. The weak lensing data is also not good enough to
justify insisting upon precisely (\ref{lensing}), so we would be
happy with $a(r) = k r b'(r)$ for any positive, order one
constant $k$. This suggests the second
MOND equation should take the form
\begin{equation}
\frac{c^2}{a_0 r^3} \Bigl( k r b' \!-\! a\Bigr)^2 = 0.
\label{MOND3}
\end{equation}
Note that it would not change the MOND phenomenology were we to
multiply (\ref{MOND3}) by a constant; we could also add a
constant times it to (\ref{MOND2}).

Relations (\ref{MOND2}-\ref{MOND3}) are the modified gravity
equations we wish to attain in the ultra-weak field regime for a
static, spherically symmetric and pressureless source. We now
seek an ultra-weak field
expansion of a Lagrangian $\mathcal{L}_{\rm MOND}$ which cancels that of general relativity $\mathcal{L}_{\rm
EH}$ and substitutes cubic terms whose variation gives
(\ref{MOND2}-\ref{MOND3}). Although one generally loses field
equations by specializing the metric before variation, we shall
recover the correct $g_{tt}$ and $g_{rr}$ equations
\cite{Palais,Torre}, in the ultra-weak field regime of course.
The equations lost by specializing first are those associated
with conservation.

After some judicious partial integrations, the Einstein-Hilbert
Lagrangian takes the form
\begin{equation}
\mathcal{L}_{\rm EH} = \frac{c^4}{16 \pi G} \, R \sqrt{-g}
\longrightarrow \Bigl(\text{Surface term}\Bigr) + \frac{c^4}{16 \pi
G} \Biggl\{-r a b' + \frac{a^2}2 + O(h^3)\Biggr\},
\end{equation}
where $h$ stands for $a$ and $b$. In the ultra-weak field regime,
the MOND Lagrangian we seek should have a quadratic term that
cancels the quadratic part of the Einstein-Hilbert Lagrangian,
plus a cubic term which enforces our interpolation (\ref{MOND2}-\ref{MOND3}) of the MOND physics. The most
general Lagrangian of this form is
\begin{equation}
\mathcal{L}_{\rm MOND} \longrightarrow \frac{c^4}{16 \pi G} \Biggl\{
\Bigl[r a b' - \frac{a^2}2 + O(h^3)\Bigr] + \frac{c^2}{a_0} \Bigl[
\frac{\alpha a^3}{r} + \beta a^2 b' + \gamma r a b^{\prime 2} +
\delta r^2 b^{\prime 3} + O(h^4)\Bigr] \Biggr\},\label{LMOND1}
\end{equation}
where $\alpha$, $\beta$, $\gamma$ and $\delta$ are dimensionless
constants whose properties we shall constrain using the
phenomenology of the ultra-weak field regime. A minor point which
deserves comment is that the $O(h^3)$ corrections to the first
square-bracketed expression in (\ref{LMOND1}) differ from the
cubic MOND terms in the second square-bracketed expression by a
factor of $a_0 r/c^2$, which would only become of order one on
horizon scales and is utterly negligible on galaxy scales.

The gravity Lagrangian is $\mathcal{L}_{\rm grav} \equiv
\mathcal{L}_{\rm EH} + \mathcal{L}_{\rm MOND}$, and we wish to
compute the variation of the associated action when specialized
to a static, spherically symmetric geometry. Recall that the full
Einstein equations for arbitrary geometry are obtained by varying
the Einstein-Hilbert action as
\begin{equation}
\frac{16 \pi G}{c^4 \sqrt{-g}} \frac{\delta S_{\rm EH}}{\delta
g^{\mu\nu}(x)} = G_{\mu\nu}(x) = \frac{8\pi G}{c^4} \, T_{\mu\nu}(x).
\end{equation}
For a static, spherically symmetric geometry $\sqrt{-g} = r^2
\sqrt{(1 \!+\! a)(1 \!+\! b)}$. We want $g_{tt} = -(1 + a)$ and
$g_{rr} = 1 + b$, and we neglect higher powers of the weak
fields, so the relevant variations for us (assuming zero pressure) are
\begin{eqnarray}
\frac{16 \pi G}{c^4 r^2} \frac{\delta S_{\rm grav}}{\delta b(r)} & =
& -\frac{c^2}{a_0 r^2} \Biggl\{ \beta (a^2)' + 2 \gamma (r a b')' +
3 \delta (r^2 b^{\prime 2})' \Biggr\} = \frac{8\pi G\rho }{c^4} \; ,
\label{gtt} \\
-\frac{16 \pi G}{c^4 r^2} \frac{\delta S_{\rm grav}}{\delta a(r)} &
= & -\frac{c^2}{a_0 r^2} \Biggl\{ \frac{3 \alpha a^2}{r} + 2 \beta a
b' + \gamma r b^{\prime 2} \Biggr\} = 0. \label{grr}
\end{eqnarray}
Demanding that equation (\ref{grr}) should have the unique
solution $a = k r b'$ implies
\begin{equation}
\alpha = -\frac{\beta}{3 k} \qquad {\rm and} \qquad \gamma = -\beta
k. \label{alphagamma}
\end{equation}
Substituting (\ref{alphagamma}) into equation (\ref{gtt}) and
demanding that it give (\ref{MOND2}) implies
\begin{equation}
\delta = -\frac16 + \frac13 \beta k^2. \label{delta}
\end{equation}
Hence the MOND Lagrangian we seek has the following expansion in
the ultra-weak field regime:
\begin{equation}
\mathcal{L}_{\rm MOND} \longrightarrow \frac{c^4 r^2}{16 \pi G}
\Biggl\{ \Bigl[\frac{a b'}{r} - \frac{a^2}{2 r^2} + O(h^3)\Bigr] +
\frac{c^2}{a_0} \Bigl[ \frac{\beta}{3 k} \Bigl(k b' -
\frac{a}{r}\Bigr)^3 - \frac{b^{\prime 3}}{6} + O(h^4)\Bigr] \Biggr\},
\label{LMOND}
\end{equation}
where the constant $\beta$ must be nonzero but is otherwise
arbitrary.

\section{Local Tools for Model Building}
\label{LOCAL}

In the previous section we considered static, spherically
symmetric geometries in the ultra-weak field limit for which MOND
ought to apply. Our result is that the form (\ref{LMOND}) for the MOND addition to the Einstein-Hilbert
Lagrangian, allows us to reproduce the MOND force and sufficient lensing without dark matter. The burden of this section is that no local,
invariant Lagrangian can have that form. Of course minus the
Einstein-Hilbert Lagrangian reproduces the quadratic parts of
(\ref{LMOND}), so it is the cubic terms which comprise the
fundamental obstacle. Our analysis is a straightforward proof by
exhaustion: we examine all local curvature scalars for static,
spherically symmetric geometries in the ultra-weak field regime.

It is useful to expand the various curvatures in powers of the
graviton field $h_{\mu\nu} \equiv g_{\mu\nu} - \eta_{\mu\nu}$. We
can greatly simplify the analysis by working in Cartesian
coordinates, which makes the affine connection vanish in the
absence of curvature. The $3+1$ decomposition of the graviton
field is
\begin{equation}
h_{00} = -b(r) \qquad , \qquad h_{0i} = 0 \qquad {\rm and} \qquad
h_{ij} = a(r) \widehat{r}^i \widehat{r}^j,
\end{equation}
where $\widehat{r}^i \equiv x^i/r$. It is also useful to introduce
the projector
\begin{equation}
\pi^{ij} \equiv \delta^{ij} - \widehat{r}^i \widehat{r}^j.
\end{equation}
The nonzero components of the affine connection are
\begin{equation}
\Gamma^0_{~0i} = \frac{b'}2 \widehat{r}^i + O(h^2) \;\; , \;\;
\Gamma^i_{~00} = \frac{b'}2 \widehat{r}^i + O(h^2) \;\; , \;\;
\Gamma^i_{~jk} = \frac{a'}2 \widehat{r}^i \widehat{r}^j
\widehat{r}^k + \frac{a}r \widehat{r}^i \pi^{jk} + O(h h').
\label{connection}
\end{equation}
In denoting higher order corrections, we make no distinction
between derivatives and inverse powers of $r$. So the term $O(h
h')$ includes terms of the form $h h'$ and $h^2/r$.

Our convention for the Riemann tensor is
\begin{equation}
R^{\rho}_{~\sigma\mu\nu} \equiv \partial_{\mu} \Gamma^{\rho}_{~\nu
\sigma} - \partial_{\nu} \Gamma^{\rho}_{~\mu \sigma} +
\Gamma^{\rho}_{~\mu \alpha} \Gamma^{\alpha}_{~\nu \sigma} -
\Gamma^{\rho}_{~\nu \alpha} \Gamma^{\alpha}_{~\mu \sigma}.
\end{equation}
Its nonzero components for a static, spherically symmetric
geometry are
\begin{eqnarray}
R_{0i0j} & = & \frac{b''}{2} \widehat{r}^i \widehat{r}^j
+ \frac{b'}{2r} \pi^{ij} + O({h'}^2), \\
R_{ijk\ell} & = & \frac{b'}{2r} \Bigl[\widehat{r}^i \widehat{r}^k
\pi^{j\ell} \!-\! \widehat{r}^{k} \widehat{r}^{j} \pi^{\ell i} \!+\!
\widehat{r}^{j} \widehat{r}^{\ell} \pi^{ik} \!-\! \widehat{r}^{\ell}
\widehat{r}^{i} \pi^{kj} \Bigr] + \frac{a}{r^2} \Bigl[\pi^{ik}
\pi^{j\ell} \!-\! \pi^{i\ell} \pi^{jk}\Bigr] + O({h'}^2). \qquad
\end{eqnarray}
We define the Ricci tensor as $R_{\mu\nu} \equiv
R^{\rho}_{~\mu\rho\nu}$, and its nonzero components in our
geometry are
\begin{eqnarray}
R_{00} & = & \frac{b''}{2} + \frac{b'}{r} + O({h'}^2), \label{R00} \\
R_{ij} & = & \Bigl[-\frac{b''}{2} \!+\! \frac{a'}{r}\Bigr]
\widehat{r}^i \widehat{r}^j + \Bigl[-\frac{b'}{2r} \!+\!
\frac{a'}{2r} \!+\! \frac{a}{r^2} \Bigr] \pi^{ij} + O({h'}^2).
\end{eqnarray}
The Ricci scalar is $R \equiv g^{\mu\nu}
R_{\mu\nu}$, and it works out to be
\begin{equation}
R = -b'' -\frac{2b'}{r} + \frac{2a'}{r} + \frac{2a}{r^2} + O({h'}^2).
\label{Ricci}
\end{equation}

From the preceding analysis we note that every nonzero component
of the curvature involves two derivatives (or inverse powers of
$r$) acting on one or more weak field,
\begin{equation}
{\rm Curvature} \sim h'' + O({h'}^2).
\end{equation}
No matter how the indices are contracted, N factors of the
curvature must therefore have the form
\begin{equation}
\Bigl( {\rm Curvature}\Bigr)^N \sim (h'')^N + O\Bigl((h')^2
(h'')^{N-1}\Bigr).
\end{equation}
The MOND correction (\ref{LMOND}) we seek involves powers of just
one derivative acting on a single weak field,
\begin{equation} \label{CUBICMONDMOND}
\mathcal{L}_{\rm MOND} \sim \frac{c^4 r^2}{16 \pi G} \Biggl\{ (h')^2
+ \frac{c^2}{a_0} (h')^3 + O(h^4) \Biggr\}.
\end{equation}
The Ricci scalar (\ref{Ricci}) gives the quadratic terms because
its linear part is a total derivative. However, the cubic terms of (\ref{CUBICMONDMOND})
not only have too few derivatives per weak field, they also
contribute an {\it odd} total number of derivatives. The latter
problem is much worse than the former because the leading weak
field term in a curvature scalar might drop out --- as it does for $R$ --- but nothing
can change the total number of the derivatives it contains.
Including differentiated curvatures increases the number of
derivatives per weak field, and can in any case only add an even
number of derivatives once all the indices are contracted to form
a scalar.

That completes the main argument of this section but it is worth
giving the nonzero components of the Einstein and Weyl tensors
for future reference:
\begin{eqnarray}
G_{00} & = & \frac{a'}{r} + \frac{a}{r^2} + O({h'}^2), \\
G_{ij} & = & \Bigl[\frac{b'}{r} - \frac{a}{r^2}\Bigr] \widehat{r}^i
\widehat{r}^j + \Bigl[\frac{b''}{2} + \frac{b'}{2 r} - \frac{a'}{2
r}\Bigr] \pi^{ij} + O({h'}^2), \\
C_{0i0j} & = & -\frac1{12} \Bigl[b'' -\frac{b'}{r} + \frac{a'}{r} -
\frac{2a}{r^2} \Bigr] \Bigl( \delta^{ij} - 3 \widehat{r}^i
\widehat{r}^j\Bigr) + O({h'}^2), \\
C_{ijk\ell} & = & -\frac16 \Bigl[b'' -\frac{b'}{r} + \frac{a'}{r} -
\frac{2a}{r^2} \Bigr] \Biggl\{ \Bigl( \delta^{ik} \!-\! \frac32
\widehat{r}^i \widehat{r}^k\Bigr) \Bigl( \delta^{j\ell} \!-\!
\frac32 \widehat{r}^j
\widehat{r}^{\ell}\Bigr) \nonumber \\
& & \hspace{5cm} - \Bigl( \delta^{i\ell} \!-\! \frac32 \widehat{r}^i
\widehat{r}^{\ell}\Bigr) \Bigl( \delta^{jk} \!-\! \frac32
\widehat{r}^j \widehat{r}^k\Bigr) \Biggr\} + O({h'}^2). \qquad
\end{eqnarray}
Note that all components of the Weyl tensor are proportional to
the same linear combination of the weak fields, so that any
scalar formed from $C_{\rho\sigma\mu\nu}$ will access this
combination
\begin{equation}
C^{\rho\sigma\mu\nu} C_{\rho\sigma\mu\nu} = \frac13 \Bigl[-b'' +
\frac{b'}{r} - \frac{a'}{r} + \frac{2 a}{r^2}\Bigr]^2 + O(h''
{h'}^2).
\end{equation}
It is also worth noting some of the other scalars we can get:
\begin{equation}
R^2 = \Bigl[-b'' - \frac{2 b'}{r} + \frac{2 a'}{r} + \frac{2 a}{r^2}
\Bigr]^2 + O(h'' {h'}^2),
\end{equation}
\begin{equation}
R^{\rho\sigma\mu\nu} R_{\rho\sigma\mu\nu} - 4 R^{\mu\nu} R_{\mu\nu}
+ R^2 = -\frac4{r^2} \Bigl(ab'' + a'b'\Bigr) + O(h'' {h'}^2).
\end{equation}

\section{Nonlocal Tools for Model Building}
\label{NONLOCAL}

The new features that nonlocality brings to model building are that
inverse differential operators reduce the number of derivatives, and
that the gradient of the invariant volume of the past light-cone allows
us to define a timelike 4-vector with which we can select particular
components of the curvature. The first feature is necessary because,
as discussed in the previous section, curvature scalars involve powers
of two derivatives (or factors of $1/r$) acting on a weak field,
whereas the MOND correction (\ref{LMOND}) we seek to realize as a
scalar involves powers of only a single derivative of a weak field.
The second property is needed to get {\it the right} weak fields.

A philosophical digression is necessary at this point. We do not
maintain that physics is nonlocal at the fundamental level; we
believe rather that nonlocality enters through quantum corrections
to the effective field equations from loops of massless gravitons.
These induce no macroscopic nonlocality in flat space background
because their interactions are suppressed by derivatives, however,
the situation is quite different when a cosmological constant is
present. It has been argued that self-interactions between elements
of the vast ensemble of infrared gravitons produced during primordial
inflation show secular growth which eventually becomes nonperturbatively
strong \cite{Tsamis:2011ep}. Nonlocal effective field equations for
cosmology have been studied \cite{TW2} as a way of abstracting these
effects to the nonperturbative regime. Our work here will apply the
very same nonlocal tools to build a model of structure formation.
Although we work on a purely phenomenological level, it might be
possible to derive a successful model from first principles using the
Schwinger-Keldysh formalism \cite{SK}. Hence our nonlocal constructions
will always be viewed as proceeding from causal evolution, based on
the notion that the universe was released in a prepared state at some
finite time. This last point is the key to being able to define a
timelike 4-vector field and it must be accepted, even if one chooses
to disregard our motivations and treat the models we propose on a purely
phenomenological level.

\subsection{The inverse scalar d'Alembertian}
\label{inversebox}

The scalar d'Alembertian is familiar to students of general
relativity,
\begin{equation}
\Box \equiv \frac1{\sqrt{-g}} \partial_{\mu} \Bigl( \sqrt{-g}
g^{\mu\nu} \partial_{\nu} \Bigr). \label{dAlem}
\end{equation}
We define the function $F(x)$ obtained by acting $\Box^{-1}$ on
any function $f(x)$ (that is, $F(x) = \Box^{-1} f$) by the
solution of the differential equation
\begin{equation}
\Box F(x) = f(x),
\end{equation}
subject to retarded boundary conditions. Specializing to the case
of a static, spherically symmetric geometry, and a source
function $f(r)$ which falls off at infinity, we obtain an
equation which can be solved by integration,
\begin{eqnarray}
\lefteqn{\frac1{r^2 \sqrt{A(r) B(r)}} \frac{d}{dr} \Biggl[
r^2 \sqrt{\frac{B(r)}{A(r)}} \frac{dF(r)}{dr} \Biggr] = f(r)}\\
& & \Longrightarrow r^2 \sqrt{\frac{B(r)}{A(r)}} \frac{dF(r)}{dr} =
\int_0^r \!\! dr' \, r^{\prime 2} \sqrt{A(r') B(r')} \, f(r') \\
& & \Longrightarrow F(r) = -\int_r^{\infty} \! \frac{dr'}{r^{\prime 2}}
\sqrt{\frac{A(r')}{B(r')}} \int_0^{r'} \!\! dr'' \, r^{\prime\prime 2}
\sqrt{A(r'') B(r'')} \, f(r''). \qquad
\end{eqnarray}

In the weak field limit for a source $f(r)$ which is already
first order we can set $A = B = 1$. We can also change the order
of integration and perform the $r'$ integration to get
\begin{eqnarray}
\frac1{\Box} f & = & -\int_r^{\infty} \! \frac{dr'}{r^{\prime 2}}
\int_0^{r'} \!\! dr'' \,r''^2\, f(r'') + O(h^2) \\
& = & -\int_0^r \!\! dr' \, r^{\prime 2} f(r') \times \frac1{r}
- \int_{r}^{\infty} \!\! dr' \, r^{\prime 2} f(r') \times \frac1{r'}
+ O(h^2) \label{boxlimit} \\
& = & -\frac1{4\pi} \int \!\! d^3x' \, \frac{f(\Vert \vec{x}' \Vert)}{
\Vert \vec{x} \!-\! \vec{x}'\Vert} + O(h^2).
\end{eqnarray}
Of course this is the usual Coulomb Green's function. We can make
similar contact with the Lienard-Wiechert potential if we regard
the system as released at some early time labeled $t=0$,
\begin{eqnarray}
\frac1{\Box} f & = & -\frac{c}{4\pi} \int_0^{\infty} \!\! dt' \!
\int \!\! d^3x' \, \frac{\delta\Bigl( c(t \!-\! t') \!-\! \Vert\vec{x}
\!-\! \vec{x}'\Vert\Bigr)}{ \Vert \vec{x} \!-\! \vec{x}'\Vert}
\times f(\Vert \vec{x}'\Vert) + O(h^2)\\
& = & -\frac1{4\pi} \int \!\! d^3x' \, \frac{\theta\Bigl( ct \!-\!
\Vert \vec{x} \!-\! \vec{x}'\Vert\Bigr) }{\Vert \vec{x}
\!-\! \vec{x}'\Vert}
\times f(\Vert \vec{x}'\Vert) + O(h^2). \label{LW}
\end{eqnarray}

The theta function in expression (\ref{LW}) is usually irrelevant
for functions $f(r)$ which fall off rapidly, and for late times
$t$. However, it plays an important role when the function $f(r)$
happens to be constant,
\begin{equation}
\frac1{\Box} f_0 = -f_0 \int_0^{ct} \!\! dr' r' + O(h^2)
= -\frac12 f_0 (ct)^2 + O(h^2).
\end{equation}
It is well to remember that even our static, spherically
symmetric systems are embedded in a larger cosmological
background which had a beginning and is even now slightly time
dependent.

\subsection{A timelike 4-vector field}
\label{4-vector}

The preceding considerations are especially important for our
second nonlocal building block: the invariant volume of the past
light-cone. Suppose $\mathcal{S}$ is the Cauchy surface on which
the initial state was released and let $\mathcal{M}$ stand for
the spacetime manifold comprising $\mathcal{S}$ and its future.
For a general metric $g_{\mu\nu}$ we define the invariant volume
of the past light-cone from the spacetime point $x^{\mu}$ as
\begin{equation}
\mathcal{V}[g](x) = \int_{\mathcal M} \!\! d^4x' \sqrt{-g(x')}
\, \theta\Bigl(-\sigma[g](x,x')\Bigr) \theta\Bigl(\mathcal{F}[g](x,x')
\Bigr). \label{invV}
\end{equation}
Here $\sigma[g](x,x')$ is the geodesic length function introduced
by DeWitt and Brehme \cite{DWB}. In expression (\ref{invV}), we note
$\theta\Bigl(\mathcal{F}[g](x,x')\Bigr)$, a functional which is the
invariant generalization of $\theta(x^0 - {x'}^0)$ needed to restrict
the integration over ${x'}^{\mu}$ to the past of $x^{\mu}$. ($\mathcal{F}$
stands for ``forward'' in this notation.) This functional is defined
as one when the extension of the geodesic between $x^{\mu}$ and
${x'}^{\mu}$ eventually intersects the initial value surface
$\mathcal{S}$, and zero otherwise.

The volume of the past light-cone is of great interest to us
because it is guaranteed to grow when the point $x^{\mu}$ evolves
in whatever is the timelike direction of the metric $g_{\mu\nu}$.
Hence its gradient must be timelike and can be used to define a
timelike vector field \cite{TW2},
\begin{equation}
u^{\mu}[g](x) \equiv -\frac{g^{\mu\nu}(x) \partial_{\nu}
\mathcal{V}[g](x)}{\sqrt{-g^{\alpha\beta}(x) \partial_{\alpha}
\mathcal{V}[g](x) \partial_{\beta} \mathcal{V}[g](x)}} \; .
\end{equation}
For the static, spherically symmetric geometry we have been
considering it reduces to
\begin{equation}
u^{\mu}[g](x) \longrightarrow \frac{\delta^{\mu}_0}{\sqrt{B(r)}}
\label{ulimit} \; .
\end{equation}
It can therefore be used to pick out the timelike components of a
tensor, just like the fundamental vector field
$\mathcal{U}_{\mu}$ of TeVeS \cite{TeVeS}.

For our purposes it is better to exploit the close relation which
exists between the volume of the past light-cone and the
functional inverse of the Paneitz operator,
\begin{equation}
D_P[g] \equiv \Box^2 + 2 D_{\mu} \Bigl[ R^{\mu\nu} - \frac13 g^{\mu\nu}
R\Bigr] D_{\nu} \; .
\end{equation}
This 4th order differential operator appears in conformal
anomalies \cite{Conf}. The relation between it and the volume of
the past light-cone is that $8\pi/D_P$ acting on one agrees with
$\mathcal{V}$ for arbitrary homogeneous and isotropic spacetimes
\cite{PW},
\begin{equation}
\frac{8\pi}{D_P[{\rm FRW}]} 1 = \mathcal{V}[{\rm FRW}].
\label{Paneitz}
\end{equation}
Perturbations away from this background do not quite agree
\cite{PW} but that is probably irrelevant for any use we might
make of $\mathcal{V}[g](x)$.

The great advantage to defining our timelike vector field using
the inverse of $D_P$ is that we can avail ourselves of a simple
partial integration trick \cite{TW1,SW1} for deriving causal and
conserved field equations. To understand the trick, consider
varying the product of a local functional of the metric $F[g]$
times some inverse differential operator ${\mathcal{D}}^{-1}$
--- either $\Box^{-1}$ or $D_P^{-1}$ --- acting on another local
functional $G[g]$,
\begin{equation}
\frac{\delta}{\delta g^{\mu\nu}(x)} \Biggl( F[g] \frac1{\mathcal{D}[g]}
G[g]\Biggr) = \frac{\delta F}{\delta g^{\mu\nu}}
\frac1{\mathcal{D}} G - F \frac1{\mathcal{D}}
\frac{\delta \mathcal{D}}{\delta g^{\mu\nu}}
\frac1{\mathcal{D}} G + F \frac1{\mathcal{D}}
\frac{\delta G}{\delta g^{\mu\nu}} \; . \label{acausal}
\end{equation}
The second and third terms on the right of expression
(\ref{acausal}) would make acausal contributions to the field
equations which involve fields to the future of $x^{\mu}$.

To see the acausality of expression (\ref{acausal}) more clearly,
let us expand out the final term, with all the implied
integrations and coordinate dependence made explicit. In order to
fix notation we express the term being varied in (\ref{acausal})
as
\begin{equation}
F \frac1{\mathcal{D}} G \equiv \int \!\! d^4x' \, F(x') \! \int \!\!
d^4x'' \, \mathcal{G}_{\rm ret}(x';x'') G(x'').
\end{equation}
Here $\mathcal{G}_{\rm ret}(x';x'')$ is the retarded Green's
function associated with the differential operator $\mathcal{D}$,
and ``retarded'' means that it vanishes for $x^{\prime\prime 0} >
x^{\prime 0}$. In this same language the final term on the right
of (\ref{acausal}) would be
\begin{equation}
F \frac1{\mathcal{D}} \frac{\delta G}{\delta g^{\mu\nu}} = \int \!\!
d^4x' \, F(x') \! \int \!\! d^4x'' \, \mathcal{G}_{\rm ret}(x';x'')
\frac{\delta G(x'')}{\delta g^{\mu\nu}(x)} \; . \label{explicit}
\end{equation}
Saying $G[g](x'')$ is local means it depends only on the metric
and some finite number of its derivatives at ${x''}^{\mu}$. Hence
its variation with respect to $g^{\mu\nu}(x)$ is proportional to
at most a finite number of derivatives of $\delta^4(x'' \!-\!
x)$. Of course this means we can perform the integration over
${x''}^{\mu}$ to get at most some derivatives acting on
$\mathcal{G}_{\rm ret}(x';x) = \mathcal{G}_{\rm adv}(x;x')$. The
remaining integration over ${x'}^{\mu}$ involves fields to the
future of $x^{\mu}$.

This sort of acausality is inevitable for any nonlocal action
based on a single field. The Schwinger-Keldysh effective field
equations avoid it by the same physical field being represented
in a complicated way with two dummy fields. One first varies with
respect to one of the dummy fields and then sets the two dummy
fields equal, after which cancellations between various
contributions result in there being no dependence upon dynamical
variables to the future of $x^{\mu}$. We shall circumvent this
complication by having recourse to the simple trick of
``partially integrating'' the acausal terms of (\ref{acausal}) so
that their nonlocality is restricted to the past of $x^{\mu}$
\cite{TW1,SW1},
\begin{eqnarray}
- F \frac1{\mathcal{D}} \frac{\delta \mathcal{D}}{\delta g^{\mu\nu}}
\frac1{\mathcal{D}} G & \longrightarrow & -
\Biggl(\frac{\delta \mathcal{D}}{\delta g^{\mu\nu}}
\frac1{\mathcal{D}} G \Biggr) \frac1{\mathcal{D}} F, \qquad \\
F \frac1{\mathcal{D}} \frac{\delta G}{\delta g^{\mu\nu}} &
\longrightarrow & \frac{\delta G}{\delta g^{\mu\nu}}
\frac1{\mathcal{D}} F. \label{trick}
\end{eqnarray}
The result is manifestly causal. It is also conserved (if we include
the variation of the measure factor) because we have just
substituted, in the field equations, the causal retarded Green's
function everywhere an acausal advanced Green's function appeared.
Conservation requires only the differential equation, which both the
advanced and retarded solutions obey. Of course this is just a
trick; a true derivation from fundamental theory would require use
of the Schwinger-Keldysh formalism \cite{SK}. However, the object of
our study is the effective field equations, and they are perfectly
valid as long as we consider them on a purely phenomenological
level.

\section{An Explicit Model}
\label{MODEL}

There are many ways to define a suitable relativistic generalization
of $\mathcal{L}_{\rm MOND}$. A particularly elegant construction is
based on two nonlocal building blocks,
\begin{eqnarray}
X[g](x) & \equiv & g^{\mu\nu} \Bigl[\partial_{\mu} \frac1{\Box} \Bigl(
R_{\alpha\beta} u^{\alpha} u^{\beta} \!-\! \frac12 R\Bigr) \Bigr]
\Bigl[\partial_{\nu} \frac1{\Box} \Bigl(R_{\rho\sigma} u^{\rho}
u^{\sigma} \!-\! \frac12 R\Bigr)\Bigr] \; \qquad \label{Xdef} \\
Y[g](x) & \equiv & g^{\mu\nu} \Bigl[\partial_{\mu} \frac1{\Box}
\Bigl(2 R_{\alpha\beta} u^{\alpha} u^{\beta} \Bigr) \Bigr]
\Bigl[\partial_{\nu} \frac1{\Box} \Bigl( 2 R_{\rho\sigma} u^{\rho}
u^{\sigma} \Bigr) \Bigr] \; . \qquad \label{Ydef}
\end{eqnarray}
Although these scalars are deeply nonlocal, even when specialized to
static and spherically symmetric geometries, they give local, and
very simple results, to lowest order in the weak field expansion. To
derive these limits recall first the weak field, static and
spherically symmetric results for $R_{00}$ and $R$ from expressions
(\ref{R00}) and (\ref{Ricci}), respectively,
\begin{eqnarray}
R_{00} & \longrightarrow & \frac1{2 r^2} \Bigl(r^2 b'\Bigr)' + O(h^2)
\; , \\
R & \longrightarrow & \frac1{r^2} \Bigl(- r^2 b' \!+\! 2 r a\Bigr)'
+ O(h^2) \; .
\end{eqnarray}
The arrow indicates specialization to static, spherically symmetric
geometries in the weak field limit. Note also that our statement
that the residues are ``$O(h^2)$'' refers only to their dependence
upon the weak fields, without regard to derivatives or powers of
$r$.

The specialization of the 4-vector $u^{\mu}[g](x)$ to a static,
spherically symmetric geometry is given by expression
(\ref{ulimit}). Hence it is just $u^{\mu}[g](x) \longrightarrow
\delta^{\mu}_0 + O(h)$ to the order we require, and we can write,
\begin{eqnarray}
R_{\alpha\beta} u^{\alpha} u^{\beta} \!-\! \frac12 R &
\longrightarrow & \frac1{r^2} \Bigl( r^2 b' \!-\! r a\Bigr)' +
O(h^2) \; , \\
2 R_{\alpha\beta} u^{\alpha} u^{\beta} & \longrightarrow &
\frac1{r^2} \Bigl( r^2 b'\Bigr)' + O(h^2) \; .
\end{eqnarray}
These terms are both first order in the weak fields so one can use
expression (\ref{boxlimit}) to implement the action of
$\square^{-1}$ on them,
\begin{eqnarray}
\frac1{\square} \Bigl(R_{\alpha\beta} u^{\alpha} u^{\beta} \!-\!
\frac12 R\Bigr) & \longrightarrow & -\int_{r}^{\infty} \!\! dr'
\Bigl( b'(r') \!-\! \frac{a(r')}{r'} \Bigr) + O(h^2) \; , \\
\frac1{\square} \Bigl( 2 R_{\alpha\beta} u^{\alpha} u^{\beta} \Bigr)
& \longrightarrow & -\int_{r}^{\infty} \!\! dr' \, b'(r') + O(h^2)
\; .
\end{eqnarray}
Of course the only derivatives that matter are with respect to the
radial coordinate $r$,
\begin{eqnarray}
\partial_{\mu} \frac1{\square} \Bigl(R_{\alpha\beta} u^{\alpha}
u^{\beta} \!-\! \frac12 R\Bigr) & \longrightarrow \delta^{r}_{\mu}
\Bigl(b' \!-\! \frac{a}{r}\Bigr) + O(h^2) \; , \label{Xsqrlim} \\
\partial_{\mu} \frac1{\square} \Bigl(2 R_{\alpha\beta} u^{\alpha}
u^{\beta}\Bigr) & \longrightarrow \delta^{r}_{\mu} b' + O(h^2) \; .
\label{Ysqrlim}
\end{eqnarray}
The residue terms in these expressions are still nonlocal. However,
substituting (\ref{Xsqrlim}) and (\ref{Ysqrlim}) into expressions
(\ref{Xdef}) and (\ref{Ydef}) gives a local result to leading order,
\begin{eqnarray}
X[g](x) & \longrightarrow & \Bigl( b' \!-\! \frac{a}{r} \Bigr)^2 +
O(h^3) \; , \label{Xlim} \\
Y[g](x) & \longrightarrow & {b'}^2 + O(h^3) \; . \label{Ylim}
\end{eqnarray}
Choosing $k = 1$ in (\ref{LMOND}), we find that acceptable MOND
equations would result from
\begin{equation}
\mathcal{L}_{\rm MOND} = \frac{c^4}{16 \pi G} \Biggl\{ \frac12
\Bigl(-X \!+\! Y\Bigr) + \frac{c^2}{6 a_0} \Bigl( |X|^{\frac32}
\!-\! |Y|^{\frac32}\Bigr) + \dots \Biggr\} \sqrt{-g} \; .
\label{weakMOND}
\end{equation}

Relation (\ref{weakMOND}) gives just the first two terms in the
ultra-weak field expansion of the theory. That is all we can
infer from the deep MONDian regime. There are many ways of
extending the expansion to all orders to define the full theory.
The chief requirement on any such extension is that it be
suitably suppressed in comparison with general relativity for
Newtonian accelerations much larger than $a_0$. Newtonian gravity
seems to be valid in the solar system out to at least 80
Astronomical Units (the furthest of the Pioneer Probes), at which
point $g_N/a_0 \sim 10^4$ \cite{far}. In this regime we can take
the Newtonian acceleration to be $g_N \sim c^2 b'(r) \sim c^2
a(r)/r$, so we need
\begin{equation}
\Bigl \vert \mathcal{L}_{\rm MOND} \Bigr\vert \ll \Bigl \vert
\mathcal{L}_{\rm GR} \Bigr\vert \sim \frac{a_0^2}{16 \pi G}
\Bigl(\frac{g_N}{a_0}\Bigr)^2, \label{suppress}
\end{equation}
for $g_N/a_0 \agt 10^4$.

It is best to study the weak-field regime using dimensionless
variables
\begin{eqnarray}
x[g] & \equiv & \frac{c^2}{3a_0} \vert X[g]\vert^{\frac12}
\longrightarrow \frac{c^2}{3a_0} \Bigl\vert b' \!-\!
\frac{a}{r}\Bigr\vert + O(h^2), \label{smallx} \\
y[g] & \equiv & \frac{c^2}{3a_0} \vert Y[g]\vert^{\frac12}
\longrightarrow \frac{c^2}{3a_0} \vert b'\vert + O(h^2).
\label{smally}
\end{eqnarray}
Whereas the variable $y[g]$ is of order one or smaller in the
ultra-weak regime, it is of order $g_N/a_0 \agt 10^4$ in the
solar system. However, the variable $x[g]$ vanishes, to lowest
order, in both regimes, so whatever function interpolates between
the two regimes must involve $y[g]$.

With the variables (\ref{smallx})--(\ref{smally}), the ultra-weak
field expansion of the MOND Lagrangian (\ref{weakMOND}) takes the
form
\begin{equation}
\mathcal{L}_{\rm MOND} = \frac{9 a_0^2}{32 \pi G} \Bigl( -x^2 + y^2
+ x^3 - y^3 + \dots \Bigr) \sqrt{-g} \; . \label{xyvars}
\end{equation}
We are therefore seeking an extension of the bracketed term in
(\ref{xyvars}) which is suppressed, relative to $y^2$, for large
$y$ and $x \sim 0$, and whose corrections to $y^3$ are
numerically small for $y\alt 1$. Of course many functions of $x$
and $y$ have this property. However, we also need to pass tests
of post-Newtonian gravity in the solar system and in binary
pulsars, therefore the suppression of (\ref{xyvars}) should be
very efficient at small distances. An extra constraint on any
possible extension of (\ref{xyvars}) is that its variation with
respect to $x$ (i.e., to the radial component of the metric, $a$)
should allow the looked-for solution $x=0$ (i.e., $a = r b'$). We
just quote here two examples of such extensions having the
required properties, and their associated behaviors for large $y$
and $x\sim 0$~:
\begin{eqnarray}
(y - x) \times \Bigl(x + x y + y\Bigr) e^{-(x + y)}
& \longrightarrow & y^2 e^{-y}, \label{example1}\\
\left(y^2 e^{-y}-x^2 e^{-x}\right)e^{-y^2} & \longrightarrow & y^2
e^{-y^2-y}. \label{example2}
\end{eqnarray}
It is easy to check that the predicted deviations from general
relativity are exponentially small with respect to the tightest
solar-system constraints, but that the MOND behavior
(\ref{xyvars}) is predicted at large distances.

The MOND Lagrangian (\ref{LMOND}) was constructed in
Sec.~\ref{PHENO} in order to cancel the general relativistic
predictions at large distances while imposing the precise physics
we wished to reproduce. In particular, we saw that it was
possible to predict any amount of weak lensing by changing the
numerical value of the coefficient $k$. In the present section,
we chose $k=1$ to recover the same weak lensing as predicted by
general relativity in presence of a dark matter halo. In such a
case, it is not necessary to cancel the $x^2$ term coming from
the Einstein-Hilbert action and to add a cubic $x^3$ as in
(\ref{xyvars}) above. Indeed, the original $x^2$ term is enough
to force $x=0$, and we may thus consider a Lagrangian depending
only on $y$, for instance
\begin{equation}
\mathcal{L}_{\rm MOND} = \frac{9 a_0^2}{32 \pi G}\;
y^2 e^{-y} \sqrt{-g} \; . \label{yAlone}
\end{equation}
Added to the Einstein-Hilbert term, this suffices to reproduce
the MOND dynamics and enough weak lensing at large distances,
while predicting fully negligible deviations from general
relativity at small distances.

None of these Lagrangians (\ref{example1}-\ref{yAlone}) is analytic
in $a_0$, but they all possess the key property of vanishing when
$a_0$ goes to zero from above. To see this, note that they vanish
for $x = 0 = y$, irrespective of $a_0$. Note also that neither $x$
nor $y$ can be negative, so if $y = k/a_0$ for some positive
constant $k$, then the limiting form, for small $a_0$, of
(\ref{example1}) and (\ref{yAlone}) vanishes like $e^{-k/a_0}$,
while the limiting form of (\ref{example2}) vanishes like
$e^{-k^2/a_0^2}$.

Our final comment on explicit models concerns the ``external field
effect'' in which MONDian behavior of one system can be severely
affected by another \cite{Wu:2007ns}. This property is deeply
embedded in the nonlocal constructions of our scalars $X[g](x)$
and $Y[g](x)$. As one can see from their definitions
(\ref{Xdef}-\ref{Ydef}), these scalars involve the nonlocal operator
$\square^{-1}$ acting on curvature scalars which are themselves
contracted into the normalized gradient $u^{\mu}[g](x)$ of the
invariant volume of the past light-cone. In the static, spherically
symmetric limit we have studied, $X[g](x)$ and $Y[g](x)$ depend only
on the central gravitating source. However, they can be quite different,
even in the static limit, when other sources are present. It is highly
significant that they also depend upon {\it past history}. This holds
out the possibility for reconciling problems in describing recently
disturbed systems such as the Bullet Cluster \cite{Dai:2008sf,Feix:2007zm}.
Of course we cannot, at this stage, claim that our model incorporates
the external field effect in a desirable way; what actually happens
beyond the static, spherically symmetric limit is a matter for future
study.

\section{Discussion}
\label{DISCUSSION}
We have considered the problem of devising a pure metric
interpolation of MOND, with neither dark matter nor additional
fields, for static, spherically symmetric systems. In the deep
MOND regime of small accelerations, gravity is described by two
weak fields, $b(r) \equiv -g_{tt} - 1$ and $a(r) \equiv g_{rr} -
1$. In this regime the MOND force law is given by equation
(\ref{MOND1}), and the requirement that there be enough weak
lensing is roughly $a(r) = k r b'(r)$ for some positive constant
$k$ of order one. Our first result is that the ultra-weak field
limiting forms of the $g_{tt}$ and $g_{rr}$ equations are
(\ref{MOND2}-\ref{MOND3}), subject only to the ambiguity of
multiplying (\ref{MOND3}) by a constant or adding such a term to
(\ref{MOND2}). Our second result is that reaching this form
requires the full gravitational Lagrangian $\mathcal{L}_{\rm
grav}$ to possess a MOND correction to the Einstein-Hilbert term,
$\mathcal{L}_{\rm grav} = \mathcal{L}_{\rm EH} + \mathcal{L}_{\rm
MOND}$, where the ultra-weak field expansion of this correction
takes the form (\ref{LMOND}).

We then turned to how the MOND Lagrangian $\mathcal{L}_{\rm
MOND}$ depends upon a general metric. Our third result is that no
local curvature scalar can reproduce the ultra-weak field form
(\ref{LMOND}). The reason is that curvature scalars involve
powers of two derivatives of a weak field whereas the MOND
correction (\ref{LMOND}) involves powers of only a single
derivative of the weak fields.

Nonlocal models have the great advantage that they allow one to
effectively remove derivatives. Our fourth result is that it is
possible to construct invariant nonlocal models which degenerate
to (\ref{LMOND}), for static and spherically symmetric geometries
in the ultra-weak field limit. In fact there seem to be many ways
to do this, some of which are laid out in section~\ref{MODEL}. So
it would be fair, at this stage, to say we are developing a class
of models rather than a unique model.

As explained in section~\ref{NONLOCAL}, our constructions involve
two nonlocal building blocks: the inverse scalar d'Alembertian
(\ref{dAlem}) and the timelike vector field $u^{\mu}[g](x)$
formed from normalizing the gradient of either the volume of the
past light-cone (\ref{invV}) or the closely related inverse of
the Paneitz operator (\ref{Paneitz}). Unlike TeVeS, the timelike
vector field of our class of models is not an independent
variable but rather a nonlocal functional of the metric itself.
In our view this nonlocality is not fundamental but should be
viewed rather as the result of quantum corrections (perhaps from
the epoch of primordial inflation) to the effective field
equations. So one should always bear in mind that our class of
models involves the universe being released in some prepared
initial state at a finite time. A derivation from fundamental
theory would be in the context of the Schwinger-Keldysh formalism
\cite{SK}. In the purely phenomenological context of our current
work, we employ the partial integration trick (\ref{trick})
introduced in \cite{TW1,SW1} to derive causal and conserved field
equations.

Although the timelike vector field $u^{\mu}[g](x)$ will certainly
introduce preferred frame effects, we believe these should only
be significant in the ultra-weak field limit for which the MOND
corrections become important. Even in this regime they should be
suppressed by the square of a peculiar velocity divided by the
speed of light. Typical peculiar velocities are several hundreds
of kilometers per second, so the suppression factor should be
about $10^{-6}$, which is not likely to be observable. However,
it may be very significant that our model depends upon events in
the past light-cone. One consequence of this dependence is that
recently disturbed systems such as the Bullet Cluster may be far
from the static MOND limit.

With any of the full metric interpolations described in
section~\ref{MODEL}, it would be possible to study the important
issues of cosmological evolution and stability. As anticipated in
\cite{SW2}, our gravitational equations (\ref{MOND2}-\ref{MOND3})
are quadratic in the ultra-weak field regime, which means the
gravitational Lagrangian is cubic. That poses an obvious potential
problem for stability, although our fears on this score might be
avoided by the absolute values needed for the fractional powers of
nonlocal scalars we employ such as (\ref{smallx}-\ref{smally}). It
should also be pointed out that the notion of energy for a nonlocal
model is subtle, and more study of this issue is certainly required.
If our class of models should prove to be unstable, it might be that
the time scale is $c/a_0 \approx 6/H_0$, which does not seem to pose
a problem for galaxy and cluster dynamics. It might even be that the
instability merely forces the weak fields back into the regime of
general relativity which is stable.

\section*{Acknowledgments}
We have benefitted from discussions and correspondence with J.
Bekenstein, S. McGaugh, M. Milgrom, R. Sanders, and C. Skordis. This
work was partially supported by European Union grant INTERREG-IIIA,
by NSF grant PHY-0855021, by the Institute for Fundamental Theory at
the University of Florida, as well as the ANR grant
``THALES''.

\end{document}